# Network Coding: Is zero error always possible?

M. Langberg
The Open University of Israel
mikel@openu.ac.il

M. Effros
California Institute of Technology
effros@caltech.edu

*Abstract*—In this work we study zero vs. $\varepsilon$-error capacity in network coding instances. For *multicast* network coding it is well known that all rates that can be delivered with arbitrarily small error probability can also be delivered with zero error probability; that is, the $\varepsilon$-error multicast capacity region and zero-error multicast capacity region are identical. For general network coding instances in which all sources originate at the same source node, Chan and Grant recently showed [ISIT 2010] that, again, $\varepsilon$-error communication has no rate advantage over zero-error communication.

We start by revisiting the setting of co-located sources, where we present an alternative proof to that given by Chan and Grant. While the new proof is based on similar core ideas, our constructive strategy complements the previous argument. We then extend our results to the setting of *index coding*, which is a special and representative form of network coding that encapsulates the "source coding with side information" problem. Finally, we consider the "edge removal" problem (recently studied by Jalali, Effros, and Ho in [Allerton 2010] and [ITA 2011]) that aims to quantify the loss in capacity associated with removing a single edge from a given network. Using our proof for co-located sources, we tie the "zero vs. $\varepsilon$-error" problem in *general* network coding instances with the "edge removal" problem. Loosely speaking, we show that the two problem are equivalent.

## I. INTRODUCTION

In the network coding paradigm, internal nodes of the network may mix the information content of the received packets before forwarding them. This mixing (or encoding) of information has been studied extensively over the last decade (see, e.g., [1], [2], [3], [4], [5] and references therein). While network coding in the *multicast* setting is well understood, far less is know about *general* network coding.

This work addresses the potential gap between *zero-error* and small non-zero error (here called "$\varepsilon$-error") communication in the context of network coding. In the multicast setting, a single source node transmits all of its information to a set of terminal nodes. In this setting, the zero- and $\varepsilon$-error capacities are the same, the capacity can be achieved precisely with linear codes, and the codes that achieve the capacity can be efficiently found [2], [3], [4]. Hence, in this setting, there is no capacity advantage in relaxing the communication requirement and enabling an $\varepsilon > 0$ error in communication.

This work was supported in part by NSF grant CCF-1018741, ISF grant 480/08 and the Open University of Israel's research fund (grant no. 46114).

For general instances of the network coding problem, in which there may be several source nodes transmitting information to different subsets of terminals, it is natural to ask whether the same phenomenon persists. When information transmitted from different sources is *dependent*, the answer is negative. That is, allowing an $\varepsilon$-error can significantly increase the achievable rate region, as shown, for example, for the Slepian-Wolf problem in [6]. In the network coding model, however, sources are assumed to be *independent*. In this case, the question of whether there is a rate advantage associated with allowing an $\varepsilon$-error remains open.

### A. Previous work

Chan and Grant explore the rate-advantage of $\varepsilon$-error communication over zero-error communication in [7]. They use the notion of *entropic* functions (e.g., [8], [9]) and their connection to the characterization of the network coding capacity [8], [9], [10] to show that $\varepsilon$-error communication has no rate advantage over zero-error communication when all sources are *co-located* at a single node. Chan and Grant also study the scenario in which the sources are not co-located, but there is a *super-node* in the network that has both full knowledge of all the information present at the sources and *low capacity* outgoing edges connecting it with each and every one of the source nodes. In this scenario, they show that $\varepsilon$-error communication again offers no rate advantage over zero-error communication [7].

### B. Our contribution

This work begins with an investigation of the relationship between zero- and $\varepsilon$-error communication in network coding. As in [7], we initially focus on networks with co-located sources. For this scenario, we present another proof that $\varepsilon$-error communication offers no rate advantage over zero-error communication. Our proof is constructive: we show how to transform any $\varepsilon$-error network code into a zero-error code at the price of a small loss in rate. As $\varepsilon$ tends to zero, the rate loss also approaches zero. Thus any rate that can be achieved with arbitrarily small error probability can also be achieved with error probability zero and arbitrarily low rate loss. The core ideas in the proof of [7] and our proof are similar; we include the proof nonetheless since the approach is central to proving the relationship between the $\varepsilon$- vs. zero-error capacity problem and the "edge removal" problem described below.

After studying co-located sources, we turn our attention to the *index coding* problem [11], which is a special instance of the network coding problem that has seen a significant amount of interest recently [11], [12], [13], [14], [15]. The index coding problem captures the problem of "source coding with side information" in which a single server wishes to communicate with several clients, each having different side information. Although index coding simulates a single-source communication problem, it does not meet the definition of either the co-located source problem or the super-node problem described above. Nevertheless, the results for co-located sources extend naturally to the index coding setting.

Finally, we consider the "edge removal" problem introduced by Jalali, Effros, and Ho in [16], [17]. Here the goal is to quantify the loss in capacity that results when a single edge is removed from a given network. While the problem is solved for a variety of special cases in [16], [17], many more cases remain unsolved. In fact, even the capacity consequences of removing edges that can carry asymptotically negligible rate are understood only in a limited family of scenarios [18], [19], [20]. Loosely speaking, we show that the "zero- vs. $\varepsilon$-error" problem in *general* network coding instances and the "edge removal" problem are equivalent. Namely, we show that quantifying the rate loss in the former problem would imply a quantification for the latter and vice-versa.

For example, as a corollary of our equivalence, we show that if removing an edge that can carry asymptotically negligible rate (that is, an edge that can carry a number of bits that grows sublinearly with the coding blocklength) has vanishing effect on the capacity of the network then $\varepsilon$-error network coding has no rate benefit over zero-error network coding. We stress that the former assumption is currently open. Our reduction between the two problems is based on our alternative proof for the "zero vs. $\varepsilon$-error" problem with co-located sources mentioned above.

The remainder of the paper is structured as follows. In Section II, we define the model of study. In Section III, we prove our reduction between zero- and $\varepsilon$-error network coding in the co-located source and super-source settings. In Section IV, we address the index coding problem. Finally, in Section V, we address the connection between the edge removal problem and the $\varepsilon$- vs. zero-error capacity problem.

## II. MODEL

An instance $\mathcal{I} = (G, S, T, B)$ of the network coding problem includes a directed acyclic network $G = (V, E)$, a set of source nodes $S \subset V$, a set of terminal nodes $T \subset V$, and an $|S|$ by $|T|$ *requirement* matrix $B$.[1] We assume, without loss of generality, that each source $s \in S$ has no incoming edges and that each terminal $t \in T$ has no outgoing edges. Let $c_e$ denote the capacity of each edge $e \in E$, and for any $k \geq 0$, define $[k]$ as $[k] = \{1, \ldots, k\}$. Then, for any block length $n$, each edge $e$ can carry one of the $2^{c_e n}$ messages in $[2^{c_e n}]$. In our setting, each source $s \in S$ holds a rate $R_s$ random variable $X_s$ uniformly distributed over $[2^{R_s n}]$. The variables of different sources are independent. A network code, $(\mathcal{F}, \mathcal{X}) = (\{f_e\} \cup \{g_t\}, \{X_e\})$ is an assignment of a pair $(X_e, f_e)$ to each edge $e \in E$, and a decoding function $\{g_t\}$ to each $t \in T$. For $e = (u, v)$, $f_e$ is a function taking as input the random variables associated with incoming edges of node $u$, and $X_e \in [2^{c_e n}]$ is the random variable equal to the evaluation of $f_e$ on its input. If $e$ is an edge leaving a source node $s \in S$, then $X_s$ is the input to $f_e$. The input to the decoding function $g_t$ consists of the random variables associated with incoming edges of terminal $t$. The output of $g_t$ is required to be a vector of all sources required by $t$. Given, the acyclic structure of $G$, the network code $(\mathcal{F}, \mathcal{X})$ can be defined by induction on the topological order of $G$.

The $|S|$ by $|T|$ requirement matrix $B = [b_{i,j}]$ has entries in the set $\{0, 1\}$, with $b_{s,t} = 1$ if and only if terminal $t$ requires information from source $s$.

A network code $(\mathcal{F}, \mathcal{X})$ is said to *satisfy* node $t$ under transmission $(\mathbf{x}_s : s \in S)$ if the decoding function $g_t$ outputs $(\mathbf{x}_s : b(s,t) = 1)$ when $(X_s : s \in S) = (\mathbf{x}_s : s \in S)$. Network code $(\mathcal{F}, \mathcal{X})$ is said to satisfy instance $\mathcal{I}$ with error probability $\varepsilon \geq 0$ if the probability that all $t \in T$ are simultaneously satisfied is at least $1 - \varepsilon$. The probability is taken over the joint distribution on random variables $(X_s : s \in S)$. Namely, $(\mathcal{F}, \mathcal{X})$ satisfies instance $\mathcal{I} = (G, S, T, B)$ with error $\varepsilon$ if

$$\Pr_{(X_s : s \in S)} [\forall \, t \in T : \, t \text{ is satisfied under } (X_s : s \in S)] \geq 1 - \varepsilon$$

An instance $\mathcal{I}$ to the network coding problem is said to be $(\varepsilon, R, n)$-feasible if there exists a network code $(\mathcal{F}, \mathcal{X})$ with block length $n$ and rate $H(X_s) = R$ (for all $s$) that satisfies $\mathcal{I}$ with error $\leq \varepsilon$. An instance $\mathcal{I}$ to the network coding problem is said to be $(\varepsilon, R)$-feasible if for any $\delta > 0$ there exists a block length $n$ such that $\mathcal{I}$ is $(\varepsilon, R(1 - \delta), n)$-feasible. Under the $\varepsilon$-error communication model, the capacity of an instance $\mathcal{I}$ refers to the supremum over all rates $R$ that are $(\varepsilon, R)$-feasible for all $\varepsilon > 0$. Often, the error probability $\varepsilon$ becomes small as the block length $n$ grows sufficiently large. Under the zero-error communication model, the capacity of an instance $\mathcal{I}$ refers to the supremum over all rates $R$ that are $(0, R)$-feasible.

Some remarks are in place. The given model assumes all sources $s \in S$ transmit information at an equal rate $R$. There is no loss of generality in this assumption as a varying rate source $s$ can be modeled by several equal rate sources all co-located at $s$.

In places throughout this work, we explicitly assume that the block length $n$ is of sufficiently large size. This is in a sense w.l.o.g. given the following claim proven in the Appendix.

---

[1] To be precise, both the set $T$ and the set $S$ should be treated as *multisets* as we allow several sources/terminals to be located at the same node.

*Claim 2.1:* Let $\mathcal{I}$ be a $(\varepsilon, R, n)$-feasible network coding instance. For any integer $c > 0$, there exists a block length $n' \geq cn$ such that $\mathcal{I}$ is also $(\varepsilon, R(1 - 5\sqrt{\varepsilon}), n')$-feasible.

## III. OUR PROOF FOR CO-LOCATED SOURCES

In this section, we consider instances $\mathcal{I}$ with co-located sources, showing that if $\mathcal{I}$ is $(\varepsilon, R)$-feasible for *any* $\varepsilon > 0$, then it is also $(0, R)$-feasible. Our proof is constructive: An arbitrary code with error probability $\varepsilon$ is used to design a zero-error code with a negligible rate loss.

*Theorem 1:* Let $\mathcal{I} = (G, S, T, B)$ be an instance to the network coding problem with $k$ sources $s_1, \ldots, s_k$ all co-located at a single vertex in $G$. If $\mathcal{I}$ is $(\varepsilon, R)$-feasible for all $\varepsilon > 0$, then it is also $(0, R)$-feasible. Specifically, for any sufficiently large block length $n$ it holds that if $\mathcal{I}$ is $(\varepsilon, R, n)$-feasible it is also

$$\left(0, R\left(1 + \frac{\log(1-\varepsilon)}{Rn} - \frac{2\log(Rn)}{Rn}\right), n\right) \text{- feasible.}$$

*Proof:* Let $(\mathcal{F}, \mathcal{X})$ be a network code of rate $R$ and block length $n$ that satisfies $\mathcal{I}$ with error probability no greater than $\varepsilon$. Then $(\mathcal{F}, \mathcal{X})$ allows the communication of source random variables $\{X_s\}_{s \in S}$, which are all independent and uniformly distributed in $[2^{Rn}]$. Let $\delta = \delta(\varepsilon)$ be a parameter to be defined later in the proof. In what follows, we show that $\mathcal{I}$ is $(0, R(1-\delta))$-feasible by constructing a new network code $(\mathcal{F}', \mathcal{X}')$ of rate $R(1-\delta)$ and the same block length $n$. Let $\{Y_s\}_{s \in S}$ denote the (new) source random variables which are all independent and uniformly distributed in $[2^{R(1-\delta)n}]$. The *new* network code $(\mathcal{F}', \mathcal{X}')$ enabling the communication of the random variables $\{Y_s\}$ uses the exact same network coding at internal nodes of the network and essentially the same decoding at terminals, the only difference is a pre-encoding step at the single source node (which holds all information on the realization of $\{Y_s\}$ to be transmitted). The ideas governing our pre-encoding are taken from the field of point to point channel coding (and especially the study of Arbitrarily Varying Channels, e.g., [21]), where it is common to find an equivalence between the notion of deterministic coding schemes with small average error and stochastic coding schemes with small maximum error.

To simplify our notation, denote the source random variables used in the original network code $(\mathcal{F}, \mathcal{X})$ by $\bar{X} = (X_1, \ldots, X_k)$ and the source random variables used in the new network code $(\mathcal{F}', \mathcal{X}')$ by $\bar{Y} = (Y_1, \ldots, Y_k)$. Denote a realization of $\bar{X} = (X_1, \ldots, X_k)$ by $\bar{\mathbf{x}} = (\mathbf{x}_1, \ldots, \mathbf{x}_k)$ and a realization of $\bar{Y} = (Y_1, \ldots, Y_k)$ by $\bar{\mathbf{y}} = (\mathbf{y}_1, \ldots, \mathbf{y}_k)$. Let $A(\bar{\mathbf{x}})$ be a function with range $\{0,1\}$ that captures the success or failure of the original communication protocol. Specifically, $A(\bar{\mathbf{x}}) = 1$ if and only if the original protocol *fails* on realization $\bar{\mathbf{x}}$ of $\bar{X}$. Here, the exact notion of "fail" is of little significance. The analysis that follows only relies on the fact for $\bar{X}$ drawn uniformly at random over $[2^{Rn}]^k$,

$$\Pr_{\bar{\mathbf{x}}}[A(\bar{\mathbf{x}}) = 1] \leq \varepsilon.$$

We now construct the pre-encoding phase of our new communication protocol $(\mathcal{F}', \mathcal{X}')$. Our pre-encoding ties the source information $\bar{\mathbf{y}}$ with a certain realization $\bar{\mathbf{x}}$ of $\bar{X}$. The new network code $(\mathcal{F}', \mathcal{X}')$ first maps the source information $\bar{\mathbf{y}}$ to its corresponding $\bar{\mathbf{x}}$. Then it proceeds using the encoding functions specified by the original network code $(\mathcal{F}, \mathcal{X})$ (with $\bar{\mathbf{x}}$ as the source information). Finally, decoding is done in two phases: first the terminals decode using the decoding functions from the original code $(\mathcal{F}, \mathcal{X})$ to obtain their relevant entries of $\bar{\mathbf{x}}$, and then they reverse the pre-encoding to obtain the corresponding entries of $\bar{\mathbf{y}}$.

Our pre-encoding is based on a "random binning" argument and is done in two steps. First, for each source $i$ we partition the set $[2^{Rn}]$ into $2^{(1-\delta)Rn}$ groups, each of size $2^{\delta Rn}$. Denote the partition for source $i$ by $\bar{P}^i = P_1^i, \ldots, P_{2^{(1-\delta)Rn}}^i$. Roughly speaking, partition $\bar{P}^i$ corresponds to random variable $Y_i$, and each realization $\mathbf{y}_i$ corresponds to a certain set $P_{\mathbf{y}_i}^i$ in $\bar{P}^i$.

Formally, to define the pre-encoding, we would like to tie each realization $\bar{\mathbf{y}} = \mathbf{y}_1, \ldots, \mathbf{y}_k$ to a certain realization $\bar{\mathbf{x}} = \mathbf{x}_1, \ldots, \mathbf{x}_k$ to be communicated over the network (using the original protocol). Each $\mathbf{x}_i$ belongs to the set corresponding to $\mathbf{y}_i$, namely $\mathbf{x}_i \in P_{\mathbf{y}_i}^i$. (Recall that we view $\mathbf{y}_i$ as an integer in $[2^{(1-\delta)Rn}]$.) The $k$ sources are encoded jointly in such a way that $A(\bar{\mathbf{x}}) = 0$ (that is, $\bar{\mathbf{x}}$ does not cause a decoding error). This ensures that if we use the original communication protocol on realization $\bar{\mathbf{x}}$ then the terminals can successfully recover the entries $\mathbf{x}_i$ that they require. Now each terminal can just check which realization $\mathbf{y}_i$ corresponds to $\mathbf{x}_i$ (that is, find the realization $\mathbf{y}_i$ such that $\mathbf{x}_i \in P_{\mathbf{y}_i}^i$) and in such a way decode $\mathbf{y}_i$.

It is left to specify how the partitions are defined and what governs our mapping between realizations $\bar{\mathbf{y}}$ and $\bar{\mathbf{x}}$. The partitions are chosen uniformly at random (and independently from each other). Now, for the mapping, consider a set of partitions $\mathbb{P} = (\bar{P}^1, \ldots, \bar{P}^k)$; partition $\bar{P}^i = (P_1^i, \ldots, P_{2^{(1-\delta)Rn}}^i)$ of alphabet $[2^{nR}]$ is used in the code for $Y_i$. We say that $\mathbb{P}$ is *good* with respect to a realization $\bar{\mathbf{y}} = (\mathbf{y}_1, \ldots, \mathbf{y}_k)$ if the product set $P_{\mathbf{y}_1}^1 \times P_{\mathbf{y}_2}^2 \times \cdots \times P_{\mathbf{y}_k}^k$ contains a realization $\bar{\mathbf{x}} = (\mathbf{x}_1, \ldots, \mathbf{x}_k)$ such that $A(\bar{\mathbf{x}}) = 0$. Indeed, if this is the case, we may map $\bar{\mathbf{y}}$ to $\bar{\mathbf{x}}$ and communicate $\bar{\mathbf{x}}$ without error over the network. It is left to show that a random set of partitions $\mathbb{P}$ is good for *all* realizations $\bar{\mathbf{y}}$ (with some positive probability). In this case we say that $\mathbb{P}$ is *good*. Lemma 3.1, below, shows that when $\delta = \delta(\varepsilon) = -\frac{\log(1-\varepsilon)}{Rn} + \frac{2\log(Rn)}{Rn}$ and $\Lambda$ is the uniform distribution over all possible partitions of the given size, $\Pr_\Lambda[\mathbb{P} \text{ is good}] \geq \frac{1}{2}$. This suffices to conclude our proof since it proves the existence of a good set $\mathbb{P}$, and the existence of a good set $\mathbb{P}$ implies a sufficient pre-encoding scheme. To prove that an instance $\mathcal{I}$ which is $(\varepsilon, R)$-feasible for all $\varepsilon > 0$ is also $(0, R)$-feasible, we may use Claim 2.1. ∎

Lemma 3.1 is an intermediate result used in the proof of Theorem 1, above. This result bounds the probability that a partition chosen uniformly at random from all partitions of the right size is "good" in the sense that every cell of the partition contains at least one element that can be decoded correctly by a given code.

*Lemma 3.1:* Let $n$ be sufficiently large. Let
$$\delta = \delta(\varepsilon) = -\frac{\log(1-\varepsilon)}{Rn} + \frac{2\log(Rn)}{Rn}.$$
Let $\Lambda$ be the uniform distribution over sets of partitions $\mathbb{P}$.
$$\Pr_{\Lambda}[\mathbb{P} \text{ is good}] \geq \frac{1}{2}.$$

*Proof:* Let the term *bad* be the complement of *good*. The proof works to show that
$$\Pr_{\Lambda}[\mathbb{P} \text{ is bad for } \bar{\mathbf{y}}] \leq \frac{1}{2} \cdot 2^{-(1-\delta)kRn}$$
for any given $\bar{\mathbf{y}} = (\mathbf{y}_1, \ldots, \mathbf{y}_k)$. We then obtain our assertion by the union bound over the $2^{(1-\delta)kRn}$ values of $\bar{\mathbf{y}}$.

Recall that the event

"$\mathbb{P}$ is bad for $\bar{\mathbf{y}}$"

is (by definition) exactly the event

"$\forall \bar{\mathbf{x}} \in P_{\mathbf{y}_1}^1 \times P_{\mathbf{y}_2}^2 \times \cdots \times P_{\mathbf{y}_k}^k : A(\bar{\mathbf{x}}) = 1$".

When $\bar{X}$ is drawn uniformly at random, $A(\bar{X}) = 1$ with probability at most $\varepsilon$. Thus, as a mental experiment, if one would assume that for random $\mathbb{P}$ the values of $\bar{\mathbf{x}} \in P_{\mathbf{y}_1}^1 \times P_{\mathbf{y}_2}^2 \times \cdots \times P_{\mathbf{y}_k}^k$ are uniformly and independently distributed, then one would have
$$\Pr_{\Lambda}[\mathbb{P} \text{ is bad for } \bar{\mathbf{y}}] \leq \varepsilon^{2^{\delta kRn}},$$
which would more than suffice for our needs. However, as the reader surely noticed, we are not in the setting of this mental experiment as there are dependencies between the different $\bar{\mathbf{x}}$ in $P_{\mathbf{y}_1}^1 \times P_{\mathbf{y}_2}^2 \times \cdots \times P_{\mathbf{y}_k}^k$. In what follows we show that we are, nevertheless, not far from this scenario.

To simplify the notation, fix $\bar{\mathbf{y}}$, and let $P^i = P_{\mathbf{y}_i}^i$ for each $i$. Note that when $\mathbb{P}$ is chosen uniformly at random, the sets $\{P^i\}_{i=1}^k$ are uniformly and independently distributed subsets of size $2^{\delta Rn}$ of $[2^{Rn}]$. Denote the $2^{\delta Rn}$ elements of $P^i$ as
$$P^i = \{\mathbf{x}_1^i, \ldots, \mathbf{x}_{2^{\delta Rn}}^i\}.$$

While the choice of any two elements $\mathbf{x}_j^i$ and $\mathbf{x}_{j'}^i$ in cell $P_i$ for source $i$ are dependent the choice of any two elements $\mathbf{x}_j^i$ and $\mathbf{x}_{j'}^{i'}$ for distinct sources $i \neq i'$ are *independent*.

To obtain our bounds, we analyze an event that has probability *greater* than the event that we want to bound. Namely, we study
$$\Pr_{\Lambda}\left[\forall j = 1, \ldots, 2^{\delta Rn} : A(\mathbf{x}_j^1, \ldots, \mathbf{x}_j^k) = 1\right]. \quad (1)$$

Notice that the above equation does not treat the probability that $A(\bar{\mathbf{x}}) = 1$ for all $(2^{\delta Rn})^k$ values of $\bar{\mathbf{x}}$ in $P^1 \times P^2 \times \cdots \times P^k$. Rather, it restricts attention to $2^{\delta Rn}$ elements $(\mathbf{x}_j^1, \ldots, \mathbf{x}_j^k) \in P^1 \times \cdots \times P^k$. (We refer to these elements as *diagonal* elements.) Since each entry $\mathbf{x}_j^i$ appears only *once* in this set, this restriction gives us the independence we need to simplify our analysis.

Recall that partition cell $P^i$ and its elements $\{\mathbf{x}_j^i\}_{j=1}^{2^{\delta Rn}}$ are random variables governed by the distribution $\Lambda$. Given any $j_0 \in [2^{\delta Rn}]$ consider any realization of the random variables $\mathbf{x}_j^i$ for all $i$ and $j \neq j_0$; denote this realization by $\mathcal{R}_{j_0}$. Claim 3.1, proved below, shows that
$$\Pr_{\Lambda}[A(\mathbf{x}_{j_0}^1, \ldots, \mathbf{x}_{j_0}^k) = 1 \mid \mathcal{R}_{j_0}] \leq (1 - 2^{-Rn(1-\delta)} + 2^{-Rn})^{-k}\varepsilon.$$

Thus, by the chain rule, the intersection (over $j$) of the events "$A(\mathbf{x}_j^1, \mathbf{x}_j^2, \ldots, \mathbf{x}_j^k) = 1$" has probability no greater than
$$\left(\frac{\varepsilon}{(1 - 2^{-Rn(1-\delta)} + 2^{-Rn})^k}\right)^{2^{\delta Rn}}.$$

We next complete the proof by showing that this value is less than $2^{-(1-\delta)kRn-1}$ for the $\delta$ defined in the lemma statement and sufficiently large $n$.

For ease of presentation, we introduce a new parameter $\alpha$ defined by $\alpha = \frac{-\log(1-\varepsilon)}{Rn}$ or equivalently $\varepsilon = 1 - 2^{-\alpha Rn}$. In what follows we assume that $n$ is sufficiently large such that $\alpha < 1/3$. Now let $\delta = \alpha + \frac{2\log(Rn)}{Rn}$ be defined as in the lemma statement.

Note that
$$\left(\frac{1 - 2^{-\alpha Rn}}{(1 - 2^{-Rn(1-\delta)} + 2^{-Rn})^k}\right)^{2^{\delta Rn}}$$
$$\leq \frac{(1 - 2^{-\alpha Rn})^{2^{\delta Rn}}}{(1 - 2^{-Rn(1-\delta)})^{k2^{\delta Rn}}}$$
$$\leq 2e(1 - 2^{-\alpha Rn})^{2^{\delta Rn}},$$
where the last inequality follows since $\delta$ is strictly less than $1/2$ and thus for sufficiently large values of $n$:
$$(1 - 2^{-Rn(1-\delta)})^{k2^{\delta Rn}} \geq (1 - 2^{-Rn(1-\delta)})^{2^{Rn(1-\delta)}} \geq 1/(2e).$$

It now suffices to show that
$$(1 - 2^{-\alpha Rn})^{2^{\delta Rn}} \leq 2^{-kRn} \leq \frac{1}{4e} \cdot 2^{-(1-\delta)kRn}.$$

The right most inequality follows from the fact that we are taking sufficiently large values of $n$ and by the fact that $\delta$ is bounded away from 1. For the left inequality, taking $\ln$, and using the fact that $\ln(1-x) \leq -x$, we have:
$$2^{\delta Rn} \ln(1 - 2^{-\alpha Rn}) \leq -\left(2^{Rn(\delta-\alpha)}\right),$$
which is less than $\ln(2^{-kRn})$ for $\delta - \alpha \geq \frac{2\log(Rn)}{Rn}$ (for sufficiently large $n$). This concludes the proof. ∎

Claim 3.1, used in the proof of Lemma 3.1, above, bounds the probability that the given code fails for the vector containing the $j_0$'th element of each partition cell $P_i$ when all other elements of $\{P_i\}_{i=1}^k$ are fixed.

*Claim 3.1:* Let $\{P^i\}_{i\in[k]} = \{P^i_{\mathbf{y}_i}\}_{i\in[k]}$ be the $k$ partition cells corresponding to an observed source vector $(\mathbf{y}_1,\ldots,\mathbf{y}_k)$. For each $i$, let $\{\mathbf{x}^i_1,\ldots,\mathbf{x}^i_{2^{\delta Rn}}\}$ denote the $2^{\delta Rn}$ elements of $P^i$. Given any $j_0 \in [2^{\delta Rn}]$, fix the realization $\mathbf{x}^i_j$ for all $(i,j) \in [k] \times ([2^{\delta Rn}] \setminus \{j_0\})$. Denote this realization by $\mathcal{R}_{j_0}$. Then

$$\Pr_\Lambda[A(\mathbf{x}^1_{j_0},\ldots,\mathbf{x}^k_{j_0}) = 1 \mid \mathcal{R}_{j_0}] \leq (1 - 2^{-Rn(1-\delta)} + 2^{-Rn})^{-k}\varepsilon$$

*Proof:* By our definition of the random variables $\{\mathbf{x}^i_j\}$, for any realization $\mathcal{R}_{j_0}$ the variable $\mathbf{x}^i_{j_0}$ is uniformly distributed in $[2^{Rn}] \setminus \{\mathbf{x}^i_j\}_{j\neq j_0}$. Thus, the vector $\bar{\mathbf{x}}_{j_0} = \mathbf{x}^1_{j_0}, \mathbf{x}^2_{j_0},\ldots,\mathbf{x}^k_{j_0}$ is uniformly distributed in a subset of $[2^{Rn}]^k$ of size $\Gamma = (2^{Rn} - 2^{\delta Rn} + 1)^k = 2^{Rnk}(1 - 2^{-Rn(1-\delta)} + 2^{-Rn})^k$. As $A(\bar{\mathbf{x}}) = 1$ only on an $\varepsilon$ fraction of $\bar{\mathbf{x}} \in [2^{Rn}]^k$, we conclude that the probability that $A(\bar{\mathbf{x}}) = 1$ when $\bar{\mathbf{x}}$ is uniform in any subset of size greater than $\Gamma$ is at most $2^{Rnk}\varepsilon/\Gamma$. This concludes our assertion. ∎

*Remark 3.1:* Essentially the same argument (of Theorem 1) can be applied to the scenario in which the sources are not co-located but there is a *super-node* that knows all of the source information $\{Y_s\}_{s\in S}$ and has links to all sources $s \in S$ with capacity which asymptotically (in $n$) tends to zero. In this case, for every $\bar{\mathbf{y}}$, the super-node computes the pre-encoding $\bar{\mathbf{x}} = (\mathbf{x}_1,\ldots,\mathbf{x}_k)$ and sends to source $i$ the location of $\mathbf{x}_i$ in $P^i_{\mathbf{y}_i}$. This information can be transmitted using capacity $\delta R$ links. Notice that by the analysis of Theorem 1, one may always take the location of $\mathbf{x}_i$ in $P^i_{\mathbf{y}_i}$ to be identical for all $i$ (as we analyzed the *diagonal* event in Equation (1)). This fact is used in Section V. As source $i$ knows $\mathbf{y}_i$, and the good $\mathbb{P}$ is known at all source nodes (and the super-node) – once each source knows the location of $\mathbf{x}_i$ in $P^i_{\mathbf{y}_i}$, it can transmit $\mathbf{x}_i$ as desired.

## IV. $\varepsilon$-ERROR VS. ZERO-ERROR FOR "INDEX CODING"

In this section we study special instances $\mathcal{I}$ to the network coding problem known as *index coding* instances. We show that for these special instances, one can prove a theorem similar to Theorem 1 even though the sources of index coding instances are not co-located.

We begin with a definition of the instances $\mathcal{I} = (G, S, T, B)$ corresponding to the index coding problem. The set $S$ is a set of $k$ sources $\{s_1,\ldots,s_k\}$. The set $T$ is a set of $k$ terminals $\{t_1,\ldots,t_k\}$. The graph $G$ consists of the vertices $\{s_1,\ldots,s_k\}$ and $\{t_1,\ldots,t_k\}$ and two special vertices $u$ and $v$. The edge set of $G$ consists of an edge $(s_i, u)$ from each source node $s_i$ to the bottleneck input node $u$, an edge $(u,v)$ which is the network bottleneck, an edge $(v,t_j)$ from the bottleneck output $v$ to each terminal node $t_j$, and a collection of *side information* edges $(s_i,t_j)$ directly from source $s_i$ to receiver $t_j$ for some subset $E_{side} \subset S \times T$ of source-terminal pairs. All edges are of capacity 1. The requirement matrix $B$ and the side information edges $E_{side}$ characterize the index coding instance. The index coding problem encapsulates the "source coding with side information" problem in which a single server wishes to communicate with several clients, each having different side information.

Now that we have defined index coding instances, it may be clear to the reader why the proof of Theorem 1 should extend to these instances as well. In index coding, encoding is done only at the bottleneck input node $u$, and node $u$ has access to all of the source information $\{X_s\}_{s\in S}$. We formalize this intuition below.

*Theorem 2:* Let $\mathcal{I}$ be an instance of the index coding problem with $k$ sources $s_1,\ldots,s_k$. If $\mathcal{I}$ is $(\varepsilon, R)$-feasible for all $\varepsilon > 0$, then $\mathcal{I}$ is also $(0, R)$-feasible. Specifically, for any sufficiently large block length $n$, it holds that if $\mathcal{I}$ is $(\varepsilon, R, n)$-feasible it is also

$$\left(0, R\left(1 + \frac{\log(1-\varepsilon)}{Rn} - \frac{2\log(Rn)}{Rn}\right), n'\right)\text{-feasible}$$

for a slightly larger block length $n'$ of size $n' = n - \log(1-\varepsilon) + 2\log(Rn)$.

*Proof:* As mentioned above, the proof follows the line of proof given in Theorem 1. To obtain a zero error network code $(\mathcal{F}', \mathcal{X}')$ from an $\varepsilon$ error code $(\mathcal{F}, \mathcal{X})$, one performs a pre-encoding step at node $u$ (which has knowledge of all source symbols) and uses the ideas specified in Remark 3.1 to allow decoding. Specifically, using the notation of Theorem 1, each source $s_i$ sends its information $\mathbf{y}_i$ on its outgoing edges. Node $u$, after receiving $\bar{\mathbf{y}}$, uses the pre-encoding procedure and obtains $\bar{\mathbf{x}}$. Using $\bar{\mathbf{x}}$ and the original network code $(\mathcal{F}, \mathcal{X})$, $u$ determines $\mathbf{z}$ the transmitted message on its outgoing edge $(u,v)$. In addition, $u$ acts as the super-node in Remark 3.1 and appends to $\mathbf{z}$ the (single) index specifying for all $i$ the location of $\mathbf{x}_i$ in $P^i_{\mathbf{y}_i}$. The fact that we are appending additional information of rate $\delta Rn = -\log(1-\varepsilon) + 2\log(Rn)$ to $\mathbf{z}$ is possible as the new network code has block length $n' = n - \log(1-\varepsilon) + 2\log(Rn)$.

For decoding, terminal $t_j$ receives the message $\mathbf{z}$, the messages $\mathbf{y}_i$ from edges $(s_i, t_j) \in E_{side}$, and the location of $\mathbf{x}_i$ in $P^i_{\mathbf{y}_i}$ for each such edge $(s_i, t_j) \in E_{side}$. Using this information, $t_j$ can reconstruct $\mathbf{x}_i$ for each edge $(s_i, t_j) \in E_{side}$ and thus use the decoding scheme of the original network code $(\mathcal{F}, \mathcal{X})$ to obtain any source information $\mathbf{x}_i$ it requires. Finally, $t_j$ can invert the pre-encoding to obtain the messages $\mathbf{y}_i$ it requires. ∎

## V. CONNECTION TO THE "EDGE REMOVAL" PROBLEM

In this section, we discuss connections between the question of zero- vs. $\varepsilon$-error network coding capacities and the question studied in [16], [17] addressing the maximum change in capacity that can result when a single edge is removed from a network. Namely, we consider the following two propositions and show that they are equivalent.

*Proposition 5.1 (Error reduction):* Let $\varepsilon = \varepsilon(n) > 0$. Let $\alpha = \alpha(n) = \frac{-\log(1-\varepsilon)}{n}$, so $\varepsilon = 1 - 2^{-\alpha n}$. Let $\varepsilon' \in [0, 1/2]$.

There exists a universal constant $c_1$ such that any instance $\mathcal{I} = (G, S, T, B)$ that is $(\varepsilon' + (1-\varepsilon')(1-2^{-\alpha n}), R, n)$-feasible is also $(\varepsilon', R - c_1\alpha, n)$-feasible.

*Proposition 5.2 (Edge removal):* Let $\mathcal{I} = (G, S, T, B)$ be an instance of the network coding problem. Let $e \in G$ be an edge of capacity $\alpha$. Let $\mathcal{I}' = (G', S, T, B)$ be the network coding instance obtained by replacing $G$ with the network $G'$ in which edge $e$ is removed. Let $\varepsilon' \in [0, 1/2]$. If $\mathcal{I}$ is $(\varepsilon', R, n)$-feasible then $\mathcal{I}'$ is $(\varepsilon', R - c_2\alpha, n)$-feasible for some universal constant $c_2$.

We note that in Proposition 5.1 a network code's error parameter $\varepsilon$ may be a function of the code's block length $n$. Thus both propositions are stated explicitly with the block length parameter $n$. In addition, Proposition 5.1 slightly generalizes the "zero- vs. $\varepsilon$-error" problem to the problem of "error-reduction," in which we seek to show that an $(\varepsilon' + (1-\varepsilon')\varepsilon, R, n)$-feasible instance is also $(\varepsilon', R - \delta, n)$ for a suitable $\delta = \delta(\varepsilon)$. Here, both $\varepsilon$ and $\varepsilon'$ are error parameters, the initial error term is expressed as $\varepsilon' + (1-\varepsilon')\varepsilon$ (which implies that the error term is always less than or equal to 1), and we seek to reduce the error from $\varepsilon' + (1-\varepsilon')\varepsilon$ to $\varepsilon'$. When $\varepsilon' = 0$, this is the familiar "zero- vs. $\varepsilon$-error" problem; we here treat the general case.

We now show that Proposition 5.1 holds if and only if Proposition 5.2 holds. Specifically we present two theorems below (one for each direction).

*Theorem 3:* Proposition 5.1 with parameter $c_1$ implies Proposition 5.2 with parameter $c_2$ equal to $c_1$.

*Proof:* Let $\mathcal{I} = (G, S, T, B)$ be an instance to the network coding problem. Let $e \in G$ be an edge (of capacity $\alpha$). Let $\mathcal{I}' = (G', S, T, B)$ be the network coding instance obtained by replacing $G$ with the network $G'$ in which the edge $e$ of capacity $\alpha$ is removed. Let $\mathcal{I}$ be $(\varepsilon', R, n)$-feasible, and consider the corresponding network code $(\mathcal{F}, \mathcal{X})$. As studied in [16], [17], consider the value $y_{\bar{\mathbf{x}}} \in [2^{\alpha n}]$ transmitted on $e$ for each and every setting of source information $\bar{\mathbf{x}} = (\mathbf{x}_1, \ldots, \mathbf{x}_k)$ that results in correct decoding. Here, as before, we take $\mathbf{x}_i \in [2^{Rn}]$. By an averaging argument, there exists a value $y \in [2^{\alpha n}]$ such that

$$\Pr_{\mathbf{x}}[y_{\bar{\mathbf{x}}} = y \mid \mathbf{x} \text{ results in correct decoding}] \geq 2^{-\alpha n}.$$

We construct a new code $(\mathcal{F}', \mathcal{X}')$ for $\mathcal{I}'$ which equals $(\mathcal{F}, \mathcal{X})$ on all functions except the functions corresponding to edges leaving $head(e)$. These changed functions use the *fixed* value $y$ as input instead of the value $X_e = y_{\bar{\mathbf{x}}}$ in the original code $(\mathcal{F}, \mathcal{X})$ for $\mathcal{I}$. As $(\mathcal{F}', \mathcal{X}')$ is identical to $(\mathcal{F}, \mathcal{X})$ when $y_{\bar{\mathbf{x}}} = y$, it holds that $(\mathcal{F}', \mathcal{X}')$ is a blocklength-$n$ code with rate $R$ and error probability at most $\varepsilon' + (1-\varepsilon')(1-2^{-\alpha n})$. Thus $\mathcal{I}'$ is $(\varepsilon' + (1-\varepsilon')(1-2^{-\alpha n}), R, n)$-feasible. If Proposition 5.1 holds, then this implies that $\mathcal{I}'$ is $(\varepsilon', R - c_1\alpha, n)$-feasible. Thus Proposition 5.2 follows, with $c_2 = c_1$. ∎

*Theorem 4:* Proposition 5.2 with parameter $c_2$ implies Proposition 5.1 with parameter $c_1 = c_2 + 1 + \frac{2\log(Rn)}{\alpha n}$

given that (in Proposition 5.1) $n$ is sufficiently large and $\alpha(n) = \frac{-\log(1-\varepsilon)}{n} < 1/3$.

*Proof:* Below, we consider the case in which $\varepsilon' = 0$. A similar analysis also holds for $\varepsilon' > 0$ (see remark located at end of proof). Let $\mathcal{I} = (G, S, T, B)$ be an instance to the network coding problem that is $(1 - 2^{-\alpha n}, R, n)$-feasible. We show that $\mathcal{I}$ is also $(0, R - c_1\alpha, n)$-feasible.

We consider 2 additional instances $\mathcal{I}_1 = (G_1, S_1, T, R)$ and $\mathcal{I}_2 = (G_2, S_2, T, R)$ similar to those considered in [7]. We start by defining the network $G_2$; network $G_1$ is then obtained from network $G_2$ by a single edge removal.

Network $G_2$ is obtained from $G$ by adding $k$ new source nodes $s'_1, \cdots, s'_k$, a new "super-node" $s$, and a relay node $r$. For each $s_i \in G$, there is a capacity-$R$ edge $(s'_i, s_i)$ from new source $s'_i$ to old source $s_i$. For each $s'_i \in G_2$, there is a capacity-$R$ edge $(s'_i, s)$ from new source $s'_i$ to the super-node $s$. There is a capacity-$\delta$ edge $(s, r)$ from the super-source $s$ to the relay $r$; this edge is the network *bottleneck* and the bottleneck capacity $\delta$ equals $\alpha + \frac{2\log(Rn)}{n}$. (Notice that the value of $\delta$ is set to satisfy the requirements in Theorem 1, as we have normalized by $R$.) Finally, the relay $r$ is connected to each source node $s_i$ by an edge $(r, s_i)$ of capacity $\delta$. The new source set $S_2$ is $\{s'_1, \ldots, s'_k\}$. For $\mathcal{I}_1$, we set $S_1 = S_2$, and remove the bottleneck edge $(s, r)$ of capacity $\delta$.

We prove the desired result by demonstrating the following properties:

(a) Instance $\mathcal{I}_2$ is $(1 - 2^{-\alpha n}, R, n)$-feasible.
(b) Instance $\mathcal{I}_2$ is also $(0, R - \alpha - \frac{2\log(Rn)}{n}), n)$-feasible.
(c) Instance $\mathcal{I}_1$ is $(0, R - \alpha - \frac{2\log(Rn)}{n} - c_2\alpha), n)$-feasible.
(d) Instance $\mathcal{I}$ is also $(0, R - c_1\alpha, n)$-feasible for $c_1 = c_2 + 1 + \frac{2\log(Rn)}{\alpha n}$.

The proof of (a) follows from our construction since $\mathcal{I}$ is $(1 - 2^{-\alpha n}, R, n)$-feasible by assumption. The proof of (b) follows by applying Theorem 1 (or more specifically Remark 3.1) to $\mathcal{I}_2$. The proof of (c) follows by removing edge $(s, r)$ from $G_2$ to obtain $G_1$, and then applying Proposition 5.2. Finally, for (d), we note that by our construction, any code $(\mathcal{F}, \mathcal{X})$ that is feasible for $\mathcal{I}_1$ is also feasible for $\mathcal{I}$.

For $\varepsilon' \in (0, 1/2]$, in (b) above we may reduce the error from $\varepsilon' + (1-\varepsilon')(1-2^{-\alpha n}) = 1 - (1-\varepsilon')2^{-\alpha n} \leq 1 - 2^{-\alpha n - 1}$ to 0 via Theorem 1 by considering $\alpha + 1/n$ instead of $\alpha$. Modifying the proof of Theorem 1 slightly, this also implies a value of $c_1 = c_2 + 1 + \frac{2\log(Rn)}{\alpha n}$ as stated in the assertion. ∎

We note that the reduction above implies in particular that:

*Corollary 5.1:* If for capacities $\alpha$ that vanish in the block length (i.e., $\alpha = o(1)$) Proposition 5.2 holds with $c_2$ such that $c_2\alpha = o(1)$ then a network coding instance $\mathcal{I}$ which is $(\varepsilon, R)$-feasible for all $\varepsilon > 0$ is also $(0, R)$-feasible.

It is interesting to point out that connections similar to those of Corollary 5.1 also exist between the edge removal problem for vanishing $\alpha$ and the *strong converse* problem studied in [18], [19], [20]. This forges an intriguing connection between the three problems.

## VI. CONCLUDING REMARKS

In this work we have studied the potential gain in allowing $\varepsilon$-error communication when compared to zero-error communication in the network coding scenario (where source information is independent). For the setting of co-located sources (and also that of index coding) we present an alternative proof to that of Chan and Grant [7], which allows us to prove an equivalence with the edge removal problem of [16], [17]. Both the capacity loss in the edge removal problem, and the potential gain in capacity when allowing an $\varepsilon > 0$ error in network communication remain open in this work. Nevertheless, our equivalence shows that there is no gain in $\varepsilon$- vs. zero-error communication if one can prove that the removal of an edge of low (vanishing) capacity has low (vanishing) effect on the communication capacity of the network at hand.

## ACKNOWLEDGMENTS

The authors would like to thank Chandra Nair for suggesting the codes used in the proof of Claim 2.1 of the Appendix.

## APPENDIX

### A. Proof of Claim 2.1

*Proof:* Roughly speaking, the proof is obtained by applying a standard argument in which one uses the original $(\varepsilon, R, n)$ coding scheme over multiple time instances combined with a carefully chosen *outer code*.

We start by setting some notation. Consider the original $(\varepsilon, R, n)$ communication protocol. Let the source random variables be $X_1, \ldots, X_k$; each $X_i$ uniform in $[2^{Rn}]$. For source realization $\bar{\mathbf{x}} = \mathbf{x}_1, \ldots, \mathbf{x}_k$, let $A(\bar{\mathbf{x}})$ be a function with range $\{0, 1\}$ that captures the success or failure of the original $(\varepsilon, R, n)$ communication protocol. Specifically, $A(\bar{\mathbf{x}}) = 1$ if and only if the original protocol *fails* on realization $\bar{\mathbf{x}}$ of $\bar{X}$. For $c' \geq c$ and a rate $R'$ to be specified shortly, we now consider an $(\varepsilon, R', c'n)$-feasible communication protocol obtained by applying the original protocol over $c'$ time instances (to obtain total block length $c'n$). Namely, let $\bar{Y} = Y_1, \ldots, Y_k$, with $Y_i$ uniform in $[2^{R'c'n}]$, be the new source information, and $\bar{\mathbf{y}} = \mathbf{y}_1, \ldots, \mathbf{y}_k$ denote its realization. Consider an encoding $C_i : [2^{R'c'n}] \to [2^{Rn}]^{c'}$ for each (new) source $i \in [k]$. For an input $\mathbf{y}_i$ to $C_i$ let $C_i(\mathbf{y}_i) = \mathbf{x}_i^{c'} = \mathbf{x}_{i,1}, \ldots, \mathbf{x}_{i,c'}$ be the encoding of realization $\mathbf{y}_i$. Here, for each pair $i, j$ it holds that $\mathbf{x}_{i,j} \in [2^{Rn}]$.

The new protocol has the following natural structure: the source input $\mathbf{y}_i$ is first encoded (at each source) to obtain $\mathbf{x}_i^{c'}$. The encoded source information $\mathbf{x}_i^{c'} = \mathbf{x}_{i,1}, \ldots, \mathbf{x}_{i,c'}$ is now transmitted over the network using the original protocol over $c'$ rounds of communication. The terminal node $t$ (after receiving the information of each and every communication round) first uses the original protocol to decode a (possibly corrupted) version $\mathbf{z}_i^{c'} = \mathbf{z}_{i,1}, \ldots, \mathbf{z}_{i,c'}$ of $\mathbf{x}_i^{c'} = \mathbf{x}_{i,1}, \ldots, \mathbf{x}_{i,c'}$; and then uses the error correcting capabilities of code $C_i$ to obtain $\mathbf{x}_i^{c'} = \mathbf{x}_{i,1}, \ldots, \mathbf{x}_{i,c'}$ and thus $\mathbf{y}_i$ (with high probability).

We now analyze the new (block length $c'n$ protocol). Recall, that for a random $\bar{\mathbf{x}} \in [2^{Rn}]^k$, it holds that $A(\bar{\mathbf{x}}) = 0$ with probability at least $1 - \varepsilon$. This implies that for random inputs $\{\mathbf{x}_i^{c'}\}_i = \{\mathbf{x}_{i,1}, \ldots, \mathbf{x}_{i,c'}\}_i$ it holds that the expected Hamming distance between $\mathbf{z}_i^{c'}$ and $\mathbf{x}_i^{c'}$ (defined above) is at most $\varepsilon c'$. Using the Chernoff bound, we conclude for random $\{\mathbf{x}_i^{c'}\}_i$ that

$$\Pr[\forall i = 1, \ldots, k : \quad \|\mathbf{x}_i^{c'} - \mathbf{z}_i^{c'}\|_H \leq 2\varepsilon c'] \geq 1 - k2^{-\Omega(\varepsilon c')}. \tag{2}$$

Now, for each $i$, consider the code $C_i$ obtained by taking any rate $r' = R'/R$ code of minimum distance $d' = 4\varepsilon c' + 1$ over the alphabet $[2^{Rn}]$ and applying an independent random permutation on each of its $c'$ coordinates.

We prove that with high probability over the random permutations defining $\{C_i\}_i$, it holds that with probability at least $1 - \varepsilon$ over the source information $\{Y_i\}$ that indeed $\|\mathbf{x}_i^{c'} - \mathbf{z}_i^{c'}\|_H \leq 2\varepsilon c'$. As we will show, the above assertion will essentially suffice to prove the claim. For the assertion, notice that for any $\bar{\mathbf{y}} = \mathbf{y}_1, \ldots, \mathbf{y}_k$ the corresponding values in $\{C_i(\mathbf{y}_i)\}_i = \{\mathbf{x}_i^{c'}\}_i = \{\mathbf{x}_{i,1}, \ldots, \mathbf{x}_{i,c'}\}_i$ are all independent and uniformly distributed in $[2^{Rn}]$. Thus, by Equation 2 the expected number (where the expectation is taken over the permutations defining the codes $\{C_i\}_i$) of source realizations $\bar{\mathbf{y}}$ for which the corresponding $\{\mathbf{x}_i^{c'}\}_i$ and $\{\mathbf{z}_i^{c'}\}_i$ satisfy $\forall i : \|\mathbf{x}_i^{c'} - \mathbf{z}_i^{c'}\|_H \leq 2\varepsilon c'$ is at least $2^{R'c'nk}(1 - k2^{-\Omega(\varepsilon c')})$. This implies the existence of a set of permutations (and correspondingly a set of codes $\{C_i\}_i$) for which the number of source realizations $\bar{\mathbf{y}}$ for which the corresponding $\{\mathbf{x}_i^{c'}\}_i$ and $\{\mathbf{z}_i^{c'}\}_i$ satisfy $\forall i : \|\mathbf{x}_i^{c'} - \mathbf{z}_i^{c'}\|_H \leq 2\varepsilon c'$ is at least $2^{R'c'nk}(1 - k2^{-\Omega(\varepsilon c')})$. Taking $c'$ large enough such that the term $k2^{-\Omega(\varepsilon c')}$ in Equation 2 is at most $\varepsilon$, and using the fact that the codes $C_i$ all have minimum distance $4\varepsilon c' + 1$, we conclude that the new protocol is indeed $(\varepsilon, R', c'n)$-feasible. It remains to specify the value of $r'$ (and thus that of $R' = Rr'$).

The rate $r'$ is set to be the highest rate for which there exist codes of block length $c'$ and minimum distance $d' = 4\varepsilon c' + 1$ over alphabets of size $2^{Rn}$. Using the Gilbert-Varshamov bound [22], [23] we can set $r' \geq 1 - H_{2^{Rn}}(4\varepsilon + 1/c') \geq 1 - H_2(4\varepsilon + 1/c') \geq 1 - 5\sqrt{\varepsilon}$ for large enough values of $c'$. Here, $H_q$ denotes the $q$-ary entropy function. ∎